\documentclass[showpacs,prl,twocolumn]{revtex4}

\usepackage{graphicx}% Include figure files
\usepackage{dcolumn}% Align table columns on decimal point
\usepackage{bm}% bold math
\usepackage{natbib}
\usepackage{amsmath}
\usepackage{appendix}
\usepackage{here}
\usepackage{color}
\usepackage{ulem}
\usepackage{epstopdf}
\usepackage{pdfpages}

\begin{document}

\title{Minimal self-contained quantum refrigeration machine based on four
    quantum dots}

\author{Davide Venturelli$^{1,2}$,  Rosario Fazio$^{1}$, and Vittorio Giovannetti$^{1}$}
\affiliation{$^{1}$NEST, Scuola Normale Superiore and Istituto Nanoscienze-CNR,  Piazza dei Cavalieri 7, I-56127 Pisa, Italy,\\
$^{2}$Quantum Artificial Intelligence Laboratory, NASA Ames Research Center,
Moffett Field CA 94035-1000.}

\date{\today}

\begin{abstract}
We present a theoretical study of an electronic quantum refrigerator based on four quantum dots arranged in a square configuration, in contact with as many thermal reservoirs. We show that the system implements the minimal mechanism for acting as a self-contained quantum refrigerator, by demonstrating heat extraction from the coldest reservoir and the cooling of the nearby quantum-dot.
\end{abstract}


\maketitle
The increasing interest in quantum thermal machines has its roots in the need to understand the relations between
thermodynamics and quantum mechanics~\cite{gemma,MARU}. The progress in this field may as well have important
applications in the control of heat transport in nano-devices ~\cite{GiazzottoREV}.
In a series of recent  works~\cite{PopescuREFR,PopescuEFF,PopescuVIRTUAL} the fundamental limits to the
dimensions of a quantum refrigerator have been found. It has been further demonstrated that these machines
could still attain  Carnot-efficiency~\cite{PopescuEFF}  thus  launching the call for  the implementation of the
smallest possible quantum refrigerator. Refs.\cite{PopescuREFR,PopescuEFF,PopescuVIRTUAL} considered
self-contained thermal machines defined as those that perform a cycle
without the supply of external work, their  action being grounded on the steady-state heat transfer from thermal reservoirs
at different temperatures.
The major difficulty in the realization~\cite{noiseCH,absorptionrefrig} of  self-contained refrigerators (SCRs)
is the engineering of the crucial three-body interaction enabling the coherent transition between a doubly excited state
in contact with a hot (H) and cold (C) reservoir, and a singly-excited state coupled to an intermediate (or
``room" - R) temperature bath.
We get around this problem by proposing an experimentally feasible implementation of a minimal SCR with semiconducting quantum dots (QDs) operating in the Coulomb blockade regime.
We are thus able to establish  a  connection between the general theory of quantum machines and the heat transport in nanoelectronics~\cite{GiazzottoREV}.

%%%%%%%%%%%%%%%%%%%%%%%%%%%%%%%%%%%%%
\begin{figure}
\begin{centering}
\includegraphics[width=.8\columnwidth
]{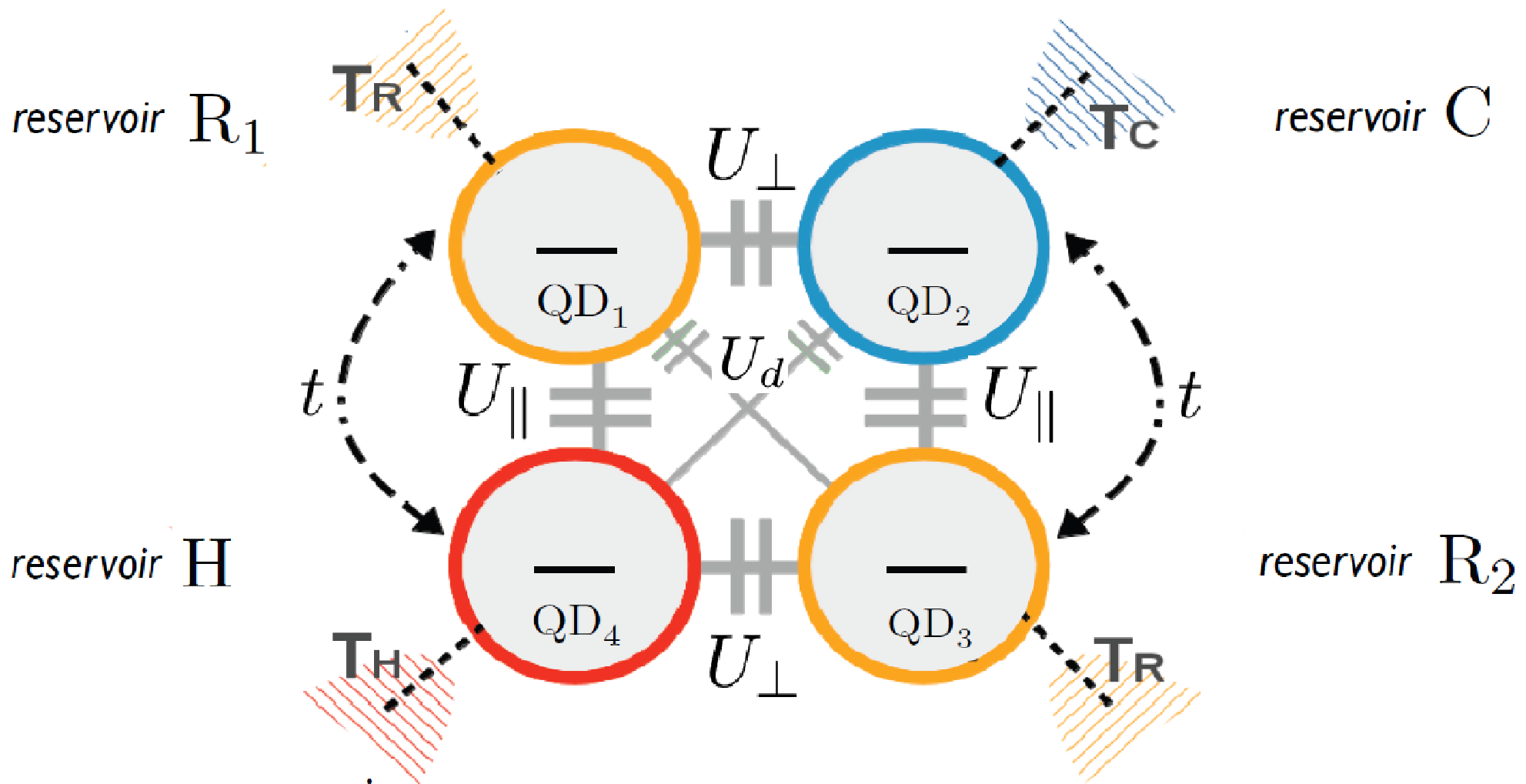}
\par\end{centering}
\caption{[Color online] The quadridot. The four quantum dots
 QD$_1$, QD$_2$, QD$_3$, and QD$_4$
 are weakly coupled to the reservoirs R$_1$, C, R$_2$, and H, respectively, which are all grounded and maintained at temperatures $T_{H}> (T_{R_1}=T_{R_2}= T_R)> T_C$. Tunneling is allowed only between  QD$_1$   and QD$_4$,  and between QD$_2$ and QD$_3$  ($t$ being the gauging parameter).
 } \label{fig:quadridot}
\end{figure}
%%%%%%%%%%%%%%%%%%%%%%%%%%%%%%%%%%%%%%%%

QDs contacted by leads were proposed as ideal  systems for achieving high
thermopower~\cite{beenakkerstaring, boese,barnasthermop} or  anomalous
thermal effects~\cite{SanchezButtiker}. Here we study a four-QD planar array (hereafter named a ``quadridot"
for simplicity) coupled to independent electron reservoirs as shown in Fig.~\ref{fig:quadridot};
with proper (but realistic) tuning of the parameters, we will show that  the quadridot acts as a SCR which pumps energy {\it{from}} the high temperature
reservoir  H and the low temperature reservoir C  {\it{to}} the intermediate temperature reservoirs R$_1$, R$_2$. Furthermore
we will analyze the conditions under which the quadridot is able to cool the dot QD$_2$ which is directly connected to the bath C, at an effective temperature that  is lower than the one  it would have had in the absence of the other reservoirs.
This will lead us to introduce an  operative definition of the local effective temperature depending on the measurement setup,
and to predict the existence of  working regimes where, for instance, the refrigeration is not accompanied by the
cooling of QD$_2$.
%In deriving our results
We start analyzing  the system Hamiltonian, identifying the conditions that allows us
to mimics the behavior of the SCR of Ref.~\cite{PopescuREFR}.

\begin{figure}
\begin{centering}
\includegraphics[width=\columnwidth]{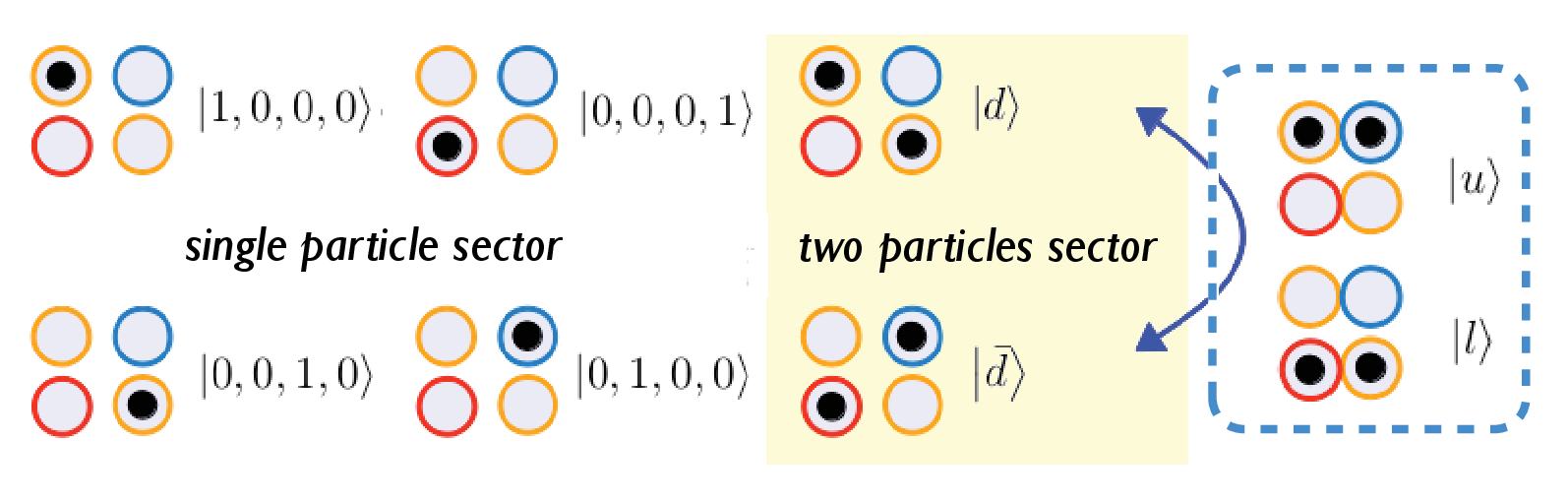}
\par\end{centering}
\caption{[Color online] Pictorial view of the low-energy electronic charged states
(a black circle indicates occupation by an electron). Due to the  hoppings terms $t$ and $g_{d \bar{d}}$  the eigenstates of the low-energy Hamiltonian are  bonding-antibonding  states $|d\rangle \pm |\bar{d}\rangle$ and the four  bonding-antibonding delocalized  single
 particles states  (the completely empty
 state  is not shown).
For $t=0$ the two electron states $|d\rangle$ and $|\bar{d}\rangle$ are resonant, while $|u\rangle$ and $|l\rangle$ are  the high-energy virtual states responsible for the  effective interaction $g_{d\bar{d}}$  coupling $|d\rangle$, $|\bar{d}\rangle$.
%The remaining  high-energy states are not drawn as their contribution to the projected dynamics is negligible.
}\label{fig:states}
\end{figure}

In the absence of the coupling to the leads, the quadridot shown in Fig. \ref{fig:quadridot} is described by the Hamitonian

\begin{equation} \nonumber
	{\cal H}_{QD} = \sum_{i=1,\cdots,4} \epsilon_i n_i + \sum_{i\neq j} \frac{U_{ij}}{2} n_i n_j - t (c^{\dagger}_1 c_4 + c^{\dagger}_2 c_3 + \mbox{h.c.}),
%\label{eq:Hdots}
\end{equation}
where for $i=1, \cdots, 4$, $c^{\dagger}_i$, $c_i$, and $n_i = c^{\dagger}_i c_i$ represent respectively
the creation, annihilation and number operators associated with the $i$-th QD. In this expression the quantities $\epsilon_i$ gauge the single particle energy levels,
% (whosevalues can be tuned by means of external gates),
$t$ defines the tunneling coupling between the dots, and
$U_{ij}$  describes the finite-range contribution of the Coulomb repulsion.
To reduce the maximum occupancy  in each QD to one electron, we will assume the
 on-site repulsion terms  $U_{ii}$ to be the largest energy scale in the problem. Furthermore, in order to mimic the dynamics of~\cite{PopescuREFR} we will take
$U_{12} = U_{34} = U_{\perp}$ and $U_{23} = U_{14} = U_{\parallel}$, both much larger than the ``diagonal" terms
$U_{24} = U_{13} = U_{d}$, and tune the
single-electron energy level of the upper-right dot (which will
be coupled to the cool reservoir C) so that $\epsilon_2 = \epsilon_1 +\epsilon_3 - \epsilon_4$.
These choices ensure that
  in the absence of tunneling ($t=0$),
   the  ``diagonal" two-particle states $|d\rangle = | 1,0,1,0\rangle$ and
$|\bar{d}\rangle = |0,1,0,1\rangle$  shown in
Fig.~\ref{fig:states}  are  degenerate (the charge states  are labeled according to the occupation of the four dots
$| n_1,n_2,n_3,n_4\rangle$). These are the only states of the
two-electron sector which play an active role in the system evolution, mimicking the
role  of the vectors $|01\rangle$, $|10\rangle$
of~\cite{PopescuREFR}. Due to the presence of $U_{\perp}$ or $U_{\parallel}$
the other  configurations are indeed much higher in energy to get permanently excited in the process. Still
 the states $|u\rangle = | 1,1,0,0\rangle$ and $|l\rangle = | 0,0,1,1\rangle$  play a fundamental role in the SCR as their
 presence  generates (via a Schrieffer-Wolff transformation~\cite{SW,SW2} and the non-zero hoppings $t$)
  an effective coupling term
between $|d\rangle$ and $|\bar{d}\rangle$ of the form $H_{eff} = g_{d \bar{d}} ( | d\rangle\langle \bar{d}| +| \bar{d}\rangle\langle d|)$ with
\begin{equation}
g_{d \bar{d}}\simeq\frac{2t^2(U_d-U_{\perp})}{(U_d-U_{\perp})^2-(\epsilon_4-\epsilon_{1})^2}\ll t. \label{eq:SWtransf}
\end{equation}
In our model $g_{d \bar{d}}$ is the analogous of the perturbative parameter $g$
of~\cite{PopescuREFR}. Its role is to
open a devoted channel
 which favors energy exchanges between the couple H-C and the couple  R$_1$-R$_2$
 by allowing  two electrons to  pass from the first to the second through the mediation of the quadridot states
  $|d\rangle$ [which is in contact with H and C] and $|\bar{d}\rangle$ [which is connected to
 R$_1$ and R$_2$].
For proper temperature imbalances this is sufficient to establish a positive heat flux {\it from} C {\it to}
QD$_2$ even if $T_C$ is the lowest of all  bath temperatures. The mechanisms can be heuristically explained as follows:  if $T_H$ is sufficiently
higher than the other bath temperatures, then the  dot  which has more chances of getting populated by its local reservoir is  QD$_4$. When
this happens, the large values of $U_\perp$, $U_\parallel$ will prevent QD$_1$ and QD$_3$ from acquiring electrons too. On the  contrary, while
QD$_4$ is populated, QD$_2$ is  allowed to accept  an electron from its reservoir C [$U_d$ being much smaller than $U_\perp$, $U_\parallel$ ]
creating $|\bar{d}\rangle$.  The coupling provided by $H_{eff}$ will then rotate the latter to  $|d\rangle$ giving the two electrons
[the one from H and the one from C] a chance of being absorbed by R$_1$-R$_2$. The opposite process (creation of $|d\rangle$ by
absorption of a couple of electrons from R$_1$-R$_2$, rotation to $|d\rangle$, and final  emission toward H-C)  is statistically suppressed
due to the (relatively) low probability that QD$_1$ or QD$_3$ will get an electron form their reservoir before QD$_4$ gets its own from H:
the net result is  a positive energy flux from H-C to R$_1$-R$_2$.

%%%%%%%%%%%%%%%%%%%%%%%
%%%%%%%%%%%%%%%%%%%%%%%%%%

To verify this picture we explicitly solve the open dynamics of the quadridot and study its asymptotic behavior. Specifically, we model our four local
baths  H, C, R$_1$ and R$_2$ as independent electron reservoirs (leads)
characterized by their own chemical potential $\mu_i$   and their own temperature [both quantities entering in   the Fermi-Dirac occupation functions $f_{i}(\epsilon)$ associated with  the reservoir]. For the sake of the simplest correspondence with the model of Ref.~\cite{PopescuREFR} in this study all $\mu_{i}$ will be set to be identical and fixed
to a  value that will be used as reference for the single particle energies of the system [e.g.  setting $\epsilon_i=0$ in the Hamiltonian corresponds to
have the i-th quadridot level at resonance with the Fermi energy of the reservoirs].
Furthermore as detailed in the caption of Fig.~\ref{fig:quadridot} the temperatures will be chosen to satisfy the relation $T_{C}<T_{R_1}=T_{R_2}<T_{H}$~\cite{NOTA}.
 Within these assumptions, the only required external action is exerted in maintaining the local equilibrium temperatures and chemical potential, in accordance to the standard definition of self-containance.
The quadridot-bath couplings (parametrized by the  amplitudes $\Gamma_{i}^{(k)}$) are hence expressed as tunneling terms of the form
\begin{equation}
H_{T}=  \; \sum_{i=1,\cdots, 4}\sum_{k} \; \Gamma_{i}^{(k)} \; c^{\dagger}_{i}a_{i,k}+\mbox{h.c.}
\label{eq:HtunRes}
\end{equation}
with $a_{i,k}$ being an annihilation operator which destroy an electron of momentum $k$ in the lead $i$.
In the Born-Markov-Secular limit~\cite{BreuerPetruccione, Schaller} these extra terms, give rise to a Lindblad equation for the reduced
density matrix  $\rho$ of the quadridot.
The presence of $t$ plus the rotation into the low-energy sector eliminates degeneracies among all possible
energy transitions between the eigenstates of the quadridot. The evolution of $\rho$ is then determined by
\begin{equation}
\dot{\rho} = \sum_{i} \mathcal{D}_{i}\left[\rho\right]\;, \label{eq:QME}
\end{equation}
where for each reservoir  $\mathcal{D}_{i}$  represents associated Lindblad dissipator.
This equation can be solved in the steady-state regime ($\dot{\rho}=0$) yielding the asymptotic configuration $\rho^\infty$,
from which   the heat currents $\langle \mathbf{J}_{Q,i}\rangle$ flowing through the i-th reservoir are then   computed as~\cite{BreuerPetruccione},
\begin{eqnarray}
\langle \mathbf{J}_{Q,i}\rangle &=& \mbox{Tr} \Big[{\cal H}_{QD} \;  \mathcal{D}_{i}\left[ \rho^\infty\right]\Big] \;. \label{eq:JQ}
\end{eqnarray}

\begin{figure}
\begin{centering}
\includegraphics[width=\columnwidth]{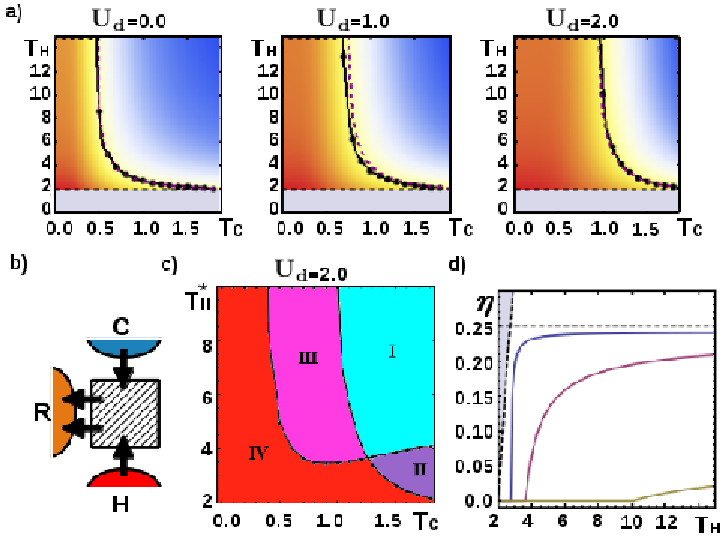}
\par\end{centering}
\caption{[Color online] a) Panels refer to $U_{d}=0$ (left), $U_{d}=1$ (center), $U_{d}=2$ (right). When $T_H$ approaches the black curve, i.e. at values very close to the analytical value of Eq.~(\ref{eq:virtualT})
(dashed line), a change of sign in the heat flow occurs. For $T_{H}$ above the threshold (blue region) the machine works as pictured in b), extracting heat from the C,H-reservoirs and pumping it into R.  In the opposite  regime (red region) heat cannot be extracted from C. Black dashed line above the grey region indicates $T_{H}=T_{R}=2$. Blue/Red background color intensity is proportional to the actual heat pumped to/extracted from C. c) Comparison between heat extraction and  single particle occupation
for $U_{d}=2$. In regions I/II the SCR is working (heat is extracted from C) while in regions III/IV the C bath \textit{receives} heat. In regions I/III we have an effective decrease of the occupation number of QD$_2$ (i.e. $\langle{n}_{2}\rangle < \langle{n}^{0}_{2}\rangle$). d) Efficiency of the SCR compared to the Carnot efficiency (dashed black line) for $T_{C}$=1.0. The curves represents $U_{d}=0$ (blue), $U_{d}=1$ (magenta) and $U_{d}=2$ (brown). The dashed horizontal line indicates maximum limit of efficiency for the quadridot  computed (for  $U_{d}=0$) as in~\cite{PopescuREFR}.}\label{fig:efficiency}
\end{figure}

If our implementation of the SCR is correct, we should see a
direct heat flow {\it{from}} the hot H and cold C reservoirs {\it{to}} the reservoirs R$_1$ and R$_2$, while the dot QD$_2$ should reach an occupation probability corresponding to an effective temperature which is lower than the one
dictated by its local reservoir C (See Fig.~\ref{fig:efficiency}-b).
 We have verified this by setting the system parameters to be consistent with those presented in~\cite{PopescuREFR} --
 making sure however that for such choice
no additional degeneracies are introduced into the system due to the larger dimension of our physical model.
While the performances of the device do not change qualitatively when varying the parameters according to above prescriptions, in the following we focus on  a specific scenario where we fixed $\epsilon_{1}=2.1$, $\epsilon_{3}=2.9$, $\epsilon_{4}=4.0$, and  $U_{\perp}=12.0$
[$U_{\parallel}$ instead is taken to be infinitely large for simplicity as its effect could be absorbed in the energy level renormalization after
the Schrieffer-Wolff transformation]. The value of $g_{d\bar{d}}$ is finally taken to be -0.001 determining
$t$ ($\lesssim 0.1$) through Eq.~(\ref{eq:SWtransf}), while the couplings
terms $\Gamma_{i}^{(k)}$ which link the quadridot to the reservoirs via  Eq.~(\ref{eq:HtunRes}) are chosen to
provide effective dissipation rates of  order~$\sim 0.0001$~\cite{NOTANEW1}.
Solving numerically the steady state equation (\ref{eq:QME}) we  observe that
for each $T_{C}<T_{R}$, there exists a minimal threshold  value
for $T_H$ above which the SCR indeed
extracts heat from the cold reservoir C. This is shown in Fig.~\ref{fig:efficiency}-a for  $T_{R}$=2 and different values of $U_{d}$,
the quadridot works as a SCR in the blue region.
Consistently with the second principle of thermodynamics the threshold value of $T_H$  (black curve in the plot) is always greater than
$T_R=2$ (for $T_H$ below $T_R$ the machine cannot produce work from H to pump heat from C),
and that the region above this threshold
 gets larger as $U_{d}$ gets smaller.  The existence of a threshold for $T_H$ implies also that, for given $T_H>T_R$,
there is a minimal temperature $T_C^*$ for the cold reservoir
under which the SCR cannot work. Interestingly for $T_H/T_R \rightarrow \infty$ the value of $T_C^*$ appears to asymptotically converge toward a finite non-zero temperature
which  depends upon the engine microscopic parameters and which can be interpreted as
{\it the emergent absolute zero}  of the model.  An approximate  analytical expression for $T_C^*$ can be derived exploiting
 the recent general theory of genuine, maximally-efficient self-contained quantum thermal machines~\cite{PopescuVIRTUAL}.
This is done by interpreting the  quadridot  as a composite system, consisting of an ``effective" virtual qubit formed
 by the states $\left|0,0,0,1\right\rangle$ and $\left|d\right\rangle$ which
 (through  $g_{d \bar d}$)
  mediates the interaction between QD$_2$ and the reservoirs H, R$_1$ and R$_2$.
 The  average occupations of the virtual qubit  levels (determined by the coupling with the reservoirs H, R$_1$ and R$_2$)
defines the effective (average)  temperature  of H, R$_1$ and R$_2$ which is perceived by QD$_2$: such temperature
competes with $T_C$ in cooling down the dot and can be identified with the value of $T_C^*$ of our model.
 Observing that  the energy levels of  $\left|0,0,0,1\right\rangle$ and $\left|d\right\rangle$ are $\epsilon_4$, $\epsilon_{1}+\epsilon_{3}+U_d$ respectively,
from~\cite{PopescuVIRTUAL} we get
 \begin{equation}
T_{C}^* \simeq T_R T_H \frac{\epsilon_{1}+\epsilon_{3}+U_d-\epsilon_4}{T_H(\epsilon_{1}+\epsilon_{3}+U_d)-T_R\epsilon_4},
\label{eq:virtualT}
\end{equation}
which fits pretty well our numerical results (See Fig.~\ref{fig:efficiency}-a) and which
for $T_H\rightarrow \infty$ yields  $T_R  (1 - {\epsilon_4}/[{\epsilon_{1}+\epsilon_{3}+U_d}])$ as emergent absolute zero of the model.

%%%%%%%%%%%%%%%%%%%%%%%
%%%%%%%%%%%%%%%%%%%%%%%%%%%%%%%%%%%%%%%%%%%%%%%

Following~\cite{PopescuEFF},
we evaluate
the ratio $\eta={\langle  \mathbf{J}_{Q,2}\rangle}/{\langle  \mathbf{J}_{Q,4}\rangle}$ between the
heat current through the cold and hot reservoirs
comparing it with
the  upper bound $({1-\frac{T_{R}}{T_{H}})/({\frac{T_{R}}{T_{C}}-1}})$ posed by the Carnot limit, and with
the  theoretical value $\eta_{th}$=$({\epsilon_{1}+\epsilon_{3}-\epsilon_4})/{\epsilon_4}$ of~\cite{PopescuREFR}
applied to the quadridot  for  $U_d=0$~\cite{NOTA1}.
The dependence of $\eta$ upon $T_H$ is plotted in Fig.~\ref{fig:efficiency}-d for different values of $U_{d}$.
We noticed that in the case  $U_d$=0 the efficiency of the quadridot converges indeed towards the theoretical value
$\eta_{th}$ of~\cite{PopescuREFR} at least for large enough $T_H$.

\paragraph{Measurements and effective local temperatures:--}
An important question is whether this refrigeration effect is accompanied with a
cooling of  QD$_2$, namely whether its effective local temperature $T_C^{(eff)}$ decreases as $T_{H}$ increases, for
sufficiently high $T_{H}$, in analogy with the qubit-cooling described in
Ref.~\cite{PopescuREFR}.
While for such idealized qubit model the definition of the local temperature is relatively straightforward, in
nanoscale systems out of equilibrium, local temperatures must be \textit{operationally} defined~\cite{TemperatureNANO}.
The most common way to proceed is
  to introduce a probe reservoir  P (a ``thermometer")  which is
 weakly coupled to that part of the system
 we are interested in
   (the dot QD$_2$ in our case) and identify the effective temperature of the latter with the value of the temperature $T_P$ of the probe
     which nullifies the heat flow through  P. This procedure yields a natural way of measuring the effect we are describing and   can be implemented easily in our model by adding an extra term
     in (\ref{eq:HtunRes}) that connects
the new reservoir P to  QD$_{2}$ with a tunnel amplitude $\Gamma_{P}$ which is much smaller than those associated with
the other reservoirs of the system
 (in the calculation we set the ratio between $\Gamma_P$ and $\Gamma_{i}$ of the other reservoirs to be of the order $10^{-3}$: this make
 sure that the presence of P does not perturb the system).
The obtained values of $T_C^{(eff)}$ are presented in Fig.~\ref{fig:efficiency}-a where it is shown that,
according to this definition of the local temperature, the conditions for cooling of QD$_{2}$ (i.e. $T_C^{(eff)}<T_C$)  are the
same for the SCR to work (implying incidentally that in this case $T_C^{(eff)}$ is always greater than the emergent zero-temperature of the system $\bar{T}_C$).

The quantity $T_C^{(eff)}$ introduced above has a clear operational meaning and according to the literature it is a good candidate to define the effective temperature of QD$_2$.
Still it is important to acknowledge   that in experiments the cooling of QD$_{2}$ can also be detected
by using the non-invasive techniques of e.g. Ref.~\cite{newexprefsQD} to look at the  decrease of
the mean asymptotic occupation number of QD$_2$, $\langle {n}_{2}\rangle =  \langle 0,1,0,0|{\rho}^{\infty}|0,1,0,0\rangle + \langle d|{\rho}^{\infty}|d\rangle$),
with respect to the same quantity  computed when the SCR is ``turned off''
(e.g.  $\langle {n}^0_{2}\rangle  =  \langle 0,1,0,0|{\rho}^{\infty}_{0}|0,1,0,0\rangle + \langle d|{\rho}^{\infty}_{0}|d\rangle$  where now
 $\rho_0^\infty$ is the asymptotic stationary state of the system reached when all the reservoirs but C are disconnected, i.e.  $\Gamma_{i\neq C} =0$).
 We notice however  that the cooling condition hereby defined does not coincide with the same pictured in Fig.~\ref{fig:efficiency}-a. We indeed exemplify in Fig.~\ref{fig:efficiency}-c for $U_{d}$=3 that according to this new definition different operating regimes are  possible for the SCR. The QD$_{2}$ might be either colder (${n}_{2} < {n}^{0}_{2}$ in zone I) or hotter (${n}_{2} > {n}^{0}_{2}$, in region II) when the device extract heat from the C reservoir. Conversely, we might achieve a colder QD$_{2}$ also when the quadridot pumps heat into the colder bath (III). In region IV none of the refrigeration effects are active. Similar regimes emerge with other activation prescriptions, such as defining $\langle {n}^0_{2}\rangle  $ as the occupation for $T_{H}=T_{R}=T_{C}$ while maintaining all tunnel couplings as constant.

\paragraph{Conclusions:--}
We conclude with experimental considerations.
Quadridots in GaAs/AlGaAs heterostructures have been implemented for Cellular-Automata computation~\cite{QCA} and  for single-electron manipulation~\cite{quadriGRENOBLE}. Strongly capacitively-coupled QDs with interdot capacitance energy ($U_{\perp}$ and $U_{\parallel}$) up to 1/3 of the intra-dot charging energy (taken to be infinite in our model) can be fabricated with current lithographic techniques \cite{gossard-strong-capac}. The diagonal inter-dot term $U_{d}$ is expected to be at most $U_{\parallel}/\sqrt{2}\simeq U_{\perp}/\sqrt{2}$ from geometrical considerations, but practically it is expected to be much smaller~\cite{quadriGRENOBLE}.
The local charging energy can be as big as 1 meV, and usually represents about 20\% of the confinement energy~\cite{klitzing-strong-capac}, which is the typical tunable values of the single-particles levels~$\epsilon_{i}$. Charging effects are expected to be further enhanced by the presence of a significant magnetic field, due to the emergence of the incompressible antidot regime in the dots~\cite{HoutenBenakkerStaring}, possibly allowing the working conditions to be achieved even more easily. In this high-field regime, the spin/orbital-Kondo effect~\cite{spinorbitalkondo,kondo4dots} is suppressed~\cite{kondomagfield}, as the transport becomes spin-polarized, so our effective description is expected to be valid.
A final ingredient for the quadridot to act as a SCR is quantum coherence.
 In QDs it is known that the main source of decoherence comes from $1/f$ noise arising from background charge fluctuations~\cite{petersson}
 (however coherent manipulation of QDs  have been reported in several experiments, e.g. see   Ref.~\cite{COHE}).
Accordingly Eq.~(\ref{eq:QME}) acquires an extra contribution whose effect (see Supplemental Material~\cite{SUPP}) is to modify
the steady state populations. In our setup as long as  the new rates are of the same order of the ones due to the coupling to the leads
 the quadridot will still work as a SCR (note, indeed, that the bounds to the blue region in Fig.~\ref{fig:efficiency} a) do not depend on these rates).
Possibly the only serious challenge is posed by the need that the induced broadening should not be too large with respect to $t$.
For the sake of simplicity we adopted small values of this parameter, however it
is  very much possible that higher values will help the efficiency of the SCR by speeding up the $|d\rangle$, $|\bar{d}\rangle$ rotations.
We finally observe that the maximum thermal energies involved should not exceed the large charging energies (i.e. $\lesssim$ 10K).

We acknowledge useful discussions with C.W.J. Beenakker, M. Carroll, F. Giazotto, F. Mazza, and J. Pekola  and support from
MIUR through the FIRB-IDEAS project RBID08B3FM, by EU through the Project IP-SOLID, and by Sandia National Laboratories.


\begin{thebibliography}{9}
\bibitem{NOTA}
Having set  $T_{R_1}=T_{R_2}=T_R$, R$_1$ and R$_2$ can be considered a single incoherent reservoir:
this hypothesis is not strictly necessary but it helps in simplifying the analysis.
\bibitem{NoteDOS}
As discussed in Ref.~\cite{SanchezButtiker}, the energy dependence of the local density of states and the tunneling amplitudes in contacted QDs alone can be the origin of interesting thermal effects. While this additional degree-of-freedom may perhaps be used to shape the working regions of the SCR, we consider for simplicity these energy dependence to be negliglibly varying over the ranges of our energy splitting.
\bibitem{NOTANEW1}
 Indicatively,  to be experimentally reasonable,  the energy units should be
considered as multiples of 100 $\mu$eV.
\bibitem{NOTA1}
Not null values of the parameter $U_{d}$ introduce  Coulomb interactions among degenerate resonant states which tend to spoil
the correspondence with the three qubit system of Ref.~\cite{PopescuREFR}.
In our case however it  is not realistic to assume $U_{d}=0$ so we keep it small but not null.
It turns out that the performances of the system are not significantly affected by this choice (e.g. see Fig.~\ref{fig:efficiency}).
\end{thebibliography}
\end{document}